# The Importance of Layer-Dependent Molecular Twisting for the Structural Anisotropy of Interfacial Water

Alexander P. Fellows[1†], Louis Lehmann[2†], Álvaro Díaz Duque[1], Martin Wolf[1], Roland R. Netz[2], and Martin Thämer[1]*

[1] Fritz-Haber-Institut der Max-Planck-Gesellschaft, Faradayweg 4-6, 14195, Berlin, Germany

[2] Department of Physics, Freie Universität Berlin, Arnimallee 14, 14195, Berlin, Germany

[†] Equal contribution

* Corresponding author

thaemer@fhi-berlin.mpg.de

(tel.): +49 (0)30 8413 5220

**Abstract**

The unique structural properties of interfacial water are at the heart of a vast range of important processes in electrochemistry, climate science, and biophysics. At interfaces, water molecules exhibit preferential orientations and an altered intermolecular H-bond connectivity. Characterising this layer-dependent anisotropic structure for such a thin molecular boundary, however, is a veritable challenge, with many important details remaining unknown. Here, we combine a novel depth-resolved second-order spectroscopy with molecular dynamics simulations to study the anisotropic structure at the air-water interface through the H-O-H bending vibration. We firstly show that the experimental nonlinear spectrum contains a large bulk-like (quadrupolar) contribution that has hampered the assessment of the interfacial structure in previous investigations. By subtracting this contribution, we uncover the elusive anisotropic interfacial response that quantitatively matches the simulated prediction. Thereafter, by analysing both the vibrational line-shape of the interfacial spectrum and its depth-dependence, we demonstrate that both the molecular tilt and twist angles of water must be highly restricted at the interface, which is confirmed by the simulated orientational distribution. Finally, by analysing the depth and orientation dependence of the bending frequency, we show substantial deviations from the expected behaviour, revealing an anomalous character to the interfacial H-bond network.

# Introduction

Understanding the anisotropic structure of water at aqueous interfaces, and particularly its evolution with depth, is of paramount importance due to its relevance in interfacial processes ranging from physiology, atmospheric chemistry, and electrochemical systems. The sheer presence of the interface induces changes to molecular orientations, diffusion dynamics, and importantly, the interconnectivity of the H-bond network.[1] These structural deviations compared to the isotropic bulk are at the heart of the aforementioned processes, however revealing their details poses a significant challenge.

A common approach for accessing the specific structural details of water is through the assessment of vibrational spectra, as the resonant line-shapes are highly sensitive to the details of the H-bond connectivity. For example, measurements of bulk water have shown that stronger bonds result in red-shifted O-H stretches and blue-shifted H-O-H bending vibrations.[2,3] In contrast to linear vibrational spectroscopies, which are dominated by these bulk signals, nonlinear optical sum-frequency generation (SFG) spectroscopy can exclusively probe the interfacial region due to its second-order selection rules. These render the responses orientationally dependent under the electric dipole approximation (EDA), leading to cancellation of signals from isotropic regions.[4–7] This unique property can make SFG a selective probe of the different aspects of the out-of-plane anisotropic structure present at the interface, namely changes in the orientational distribution (preferential orientations) and intermolecular connectivity. Furthermore, provided the phase of the nonlinear response is measured (by phase-resolved SFG), the absolute orientations can also be directly assessed.[8] Importantly, extracting the anisotropic contribution to the interfacial structure is an essential step for understanding the behaviour of the interface because it is precisely this structural perturbation from the bulk that defines its non-bulk-like properties. Gaining a clear picture of the anisotropic structure at the interface and especially its depth-dependence is, however, not straightforward simply from analysing SFG spectra. Nevertheless, it can be facilitated by comparing the experimental results with SFG spectra calculated from molecular dynamics (MD) simulations, which can directly yield the different structural motifs present at the interface and can correlate them to the observed resonant features.[9–15]

Many SFG investigations have analysed the O-H stretching modes of aqueous interfaces, with simulations reproducing the main observed spectral features, including their absolute amplitudes, but still showing slight deviations in the spectral line-shape.[13,14,16,17] From such studies at the pure air-water interface, the presence of 'free' O-H species (water molecules with 'dangling' bonds) has been identified. Furthermore, both positive and negative spectral contributions at different frequencies are observed, which are typically attributed to 'pointing up' and 'pointing down' structural motifs with differing H-bond interconnectivity, respectively. Such a description of the molecular directionality is commonly referring to the projection of the molecular dipole onto the surface normal (defining its molecular tilt angle). Overall, this paints a picture of the air-water interface where the top-most molecules are pointing towards air with dangling bonds, followed by molecules in deeper layers preferentially pointing 'down' with stronger H-bonding.

Despite these observations, a thorough interpretation of the O-H stretching responses in terms of the molecular structure is difficult. This is due to the presence of two distinct stretching modes on each molecule, either seen as a symmetric and anti-symmetric mode or as two decoupled O-H oscillators, with differing directionality. This, combined with the intrinsically different frequencies of these two modes, muddies the connection between specific spectral features and the molecular and intermolecular structure. Additional complexity then originates from the significant inter- and intra-molecular coupling and delocalisation of the stretching modes[18,19], meaning the response cannot trivially be connected to specific molecular motifs. Probing the H-O-H bending vibration is therefore often considered a more favourable approach for elucidating the molecular structure at the interface as the bending vibration is a single mode largely localised to each molecule that approximately aligns with the molecular dipole and has minimal contributions from coupling.[20–22] This hence presents a more direct and simpler route to elucidate the molecular structure at the air-water interface.

A very detailed view of the anisotropic structure at the air-water interface can thus be gained by analysing the depth/layer-dependent anisotropic response of the bending mode that is accessible from the depth-dependent second-order susceptibility, $\chi^{(2)}(z)$. This, however, requires the spectra to be of purely electric dipolar origin (i.e., containing only the interfacial dipolar signal, ID). While this is readily obtained for the calculated spectra from simulations, the extent to which the experimental SFG signals are solely of electric dipolar origin is less clear and a long-standing question in second-order spectroscopy.[23,24] Beyond the EDA, electric quadrupolar and magnetic dipolar signals (combined here into a generalized quadrupolar contribution, as per the conventions of Morita[4]) can also contribute to the obtained spectra. Unlike electric dipolar responses, these signals are insensitive to molecular orientation and thus do not report on the structural anisotropy. Therefore, the presence of any non-negligible quadrupolar signals can obscure the desired structural information. It is thus critical that the presence of these signals is either ruled out, or their contributions removed.

In this work we probe the water bending mode using our recently developed spectroscopic tool which combines phase-resolved SFG with its analogous and simultaneously generated difference-frequency response (DFG, see schematic in Figure 1a).[25,26] Due to the characteristic signatures of anisotropic (dipolar) and isotropic (quadrupolar) signals in the SFG and DFG responses, this technique can accurately separate them.[27] The results are therefore divided into two parts. In the first part, we apply the SFG-DFG technique to the water bending mode spectra of both the pure air-water interface as well as those with surface charges to isolate the ID response. This interfacial spectrum is then quantitatively compared to the prediction from MD simulations. Thereafter, in the second part, we analyse the details of its line-shape and depth-dependence. This analysis reveals that the conventional model for interpreting SFG spectra in terms of the preferential molecular orientation ('pointing up/down' molecular dipoles) is utterly insufficient for the air-water interface. Through a revised model, we show that the anisotropic water structure is not only defined by a restricted and layer-dependent molecular tilt angle, but also by a highly restricted and correlated molecular twist. These findings provide a new and more detailed picture of the anisotropic structure at the air-water interface and, by further analysing the depth-dependent frequency shifts from these structural motifs, uncover an anomalous character to the interfacial H-bonding.

## Results and Discussion
### *Obtaining the Interfacial Dipolar Spectrum*
Although traditionally assumed, the dominant role of the ID contribution in SFG spectra has recently been questioned for the case of the water bending mode at the air-water interface. Clear evidence for this comes from the striking disagreement between the measured total and calculated ID-only line-shapes, as shown in Figure 1b (only showing the imaginary parts i.e., the absorptive line-shapes). Specifically, the simulated ID-only spectrum shows a 'dip-peak' line-shape whereas the experimentally obtained response shows only a single positive resonance.[21,28,29] While initial SFG intensity measurements seemed to reproduce the simulated bipolar line-shape, this was subsequently shown to source from interference with the significant non-resonant contribution.[28–31] More accurate phase-sensitive measurements (which are not subjected to such interference effects) instead reveal the single positive resonance at ~1650 cm$^{-1}$ (Figure 1b).[32,33] This substantial discrepancy indicates that either the experimental spectrum contains significant quadrupolar contributions, or that the simulations are inaccurate, and with it potentially also our current view of the interfacial water structure altogether.

The first experimental indications for the significance of quadrupolar contributions to the bending mode spectrum came from Tahara and co-workers who performed the initial phase-sensitive measurement of the air-water interface in this frequency region.[32] Their results from studies both with and without the addition of salt indicated that the apparent single positive resonance is dominated by an interfacial quadrupolar (IQ) signal. A more rigorous analysis was later performed on charged aqueous interfaces, where there is significant field-induced molecular alignment and thus an enhanced ID response which would take the form of a single resonant band that flips sign upon charge inversion, unlike any quadrupolar signals. While different initial measurements yielded contradicting results and produced significant controversy about the mechanistic origin of the observed bending signals[31,34–36], Bakker and co-workers[33] recently demonstrated that both a charge-independent and charge-dependent contribution must be present in the measured spectra. This result confirms the significance of quadrupolar signals in the bending region, but clearly also demonstrates that ID signals can become relevant when preferential molecular orientation is induced by surface charges. This hence offers an explanation for the substantial disagreement between the experimental and simulated spectra for the pure air-water interface. Nevertheless, the experimental ID spectrum from the air-water interface that is required for a reliable structural analysis remains elusive.

The exact origin of the SFG response can be determined using the SFG-DFG technique.[25,26,37,38] SFG and DFG contributions that originate from structural anisotropy (dipolar signals) have equal intrinsic (local) responses. However, due to the different wavevector mismatches for SFG and DFG (opposite signs and different amplitudes), integration of the intrinsic (local) responses over depth leads to different phase-shifts and amplitude scaling factors. The induced phase difference between the obtained SFG and DFG spectra amounts to approximately 2° per nm in depth, making the technique a very sensitive tool for analysing the thickness of the structural anisotropy.[37] For signal contributions from structurally isotropic regions (quadrupolar) the situation is different. Here the corresponding intrinsic (local) responses in SFG and DFG are naturally opposite in sign (180° phase-shifted) and the entire isotropic bulk contributes to the signals. In consequence, the integration-induced additional phase-shift (+90 deg. for SFG and -90 deg. for DFG) leads to a vanishing phase difference but leaves the characteristic amplitude scaling (see schematic phase diagram in Figure 1c). Based on these characteristics, combining the phase-difference and amplitude ratio of the SFG and DFG spectra allows for the separation of any anisotropic and isotropic contribution.

The SFG and DFG bending responses (imaginary parts) of the pure air-water interface are depicted in Figure 1d. By comparing the two spectra, it is firstly clear that they exhibit little if any phase-difference, which would present as oppositely distorted spectral line-shapes, since both exhibit a single positive resonance with an almost symmetric line-shape centred at ~1656 cm$^{-1}$, as previously reported[33]. This observation aligns with previous work showing that the anisotropic structure at the air-water interface is contained within an exceptionally small region of ~6-8Å.[38] On further inspection, however, it is also clear that the DFG response has a larger amplitude. This difference is emphasised in the spectral integral traces also shown in Figure 1d. The observation of little phase difference but a distinct amplitude offset is therefore indicative of a large isotropic quadrupolar contribution (as schematically shown in Figure 1c)[38], clearly confirming the significance of quadrupolar contributions to the bending mode, as previously suggested. Furthermore, with the amplitude difference between SFG and DFG only arising from the integration of these quadrupolar responses over the entire probed depth, this observation incontrovertibly shows that the quadrupolar contribution must have a significant bulk-like character.

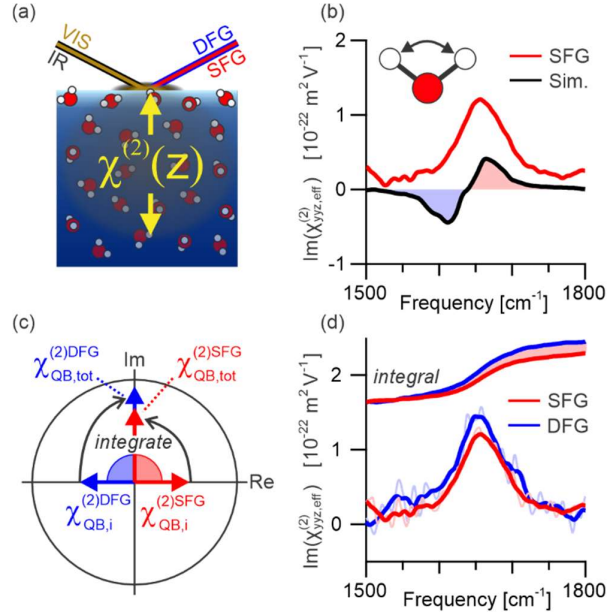

*Figure 1: (a) Schematic of depth-resolved second-order spectroscopy from the air-water interface showing the IR+visible → SFG + DFG. (b) Comparison of the experimental (SFG) spectrum of the H-O-H bending mode (red) with the ID-only spectrum calculated from MD simulations (black). (c) Schematic Argand diagram of the isotropic quadrupolar responses, showing both the intrinsic susceptibilities, $\chi^{(2)}_{QB,i}$, and their depth-integrated responses, $\chi^{(2)}_{QB,tot}$. (d) SFG (red) and DFG (blue) spectra of the H-O-H bending mode of the pure air-water interface, shown both as the obtained spectra (lighter colours) and with smoothed line-shapes. Also shown are the cumulative spectral integration traces.*

The presence of this large bulk-like contribution to the pure air-water interface spectrum clearly hinders a detailed analysis of the anisotropic interfacial water structure at this point because it buries the desired interfacial contribution. This raises the important question how large the dipolar response is and can it be retrieved from the total line-shape? To address these questions, we move to charged aqueous interfaces. In such systems, the net surface charge generates a static electric field which propagates away from the interface and creates a torque on the molecular dipoles, inducing anisotropy through preferential molecular orientation.[37,39–42] Using the Gouy-Chapman-Stern (GCS) model of these systems (shown schematically in Figure 2a), the anisotropic structure is then described by two distinct layer contributions: the compact layer (CL), which represents the thin layer of anisotropy induced by the presence of the interface, and the diffuse layer (DL), which only gives rise to a field-induced response, but can extend tens of nanometres away from the interface in low salinity conditions.[43,44] Adding charges to the surface thus has several important consequences. Firstly, the presence of the field-induced contribution generates a significantly larger dipolar response than for pure air-water but, as the isotropic bulk is essentially unperturbed by the changes at the interface, the quadrupolar contribution is left unchanged. Furthermore, since inversion of the charge also inverts the electric field and thus the direction of the molecular reorientation, a dipolar response should flip sign upon charge inversion while quadrupolar responses should not. This gives a clear route to determine the significance of both contributions to the overall signal. Finally, as the CL and DL contributions to the dipolar signal arise from different depths, SFG-DFG can fully resolve their contributions and isolate their spectral line-shapes.[37] Therefore, since the DL contribution is essentially 'bulk-like',

SFG-DFG can separate and obtain the bulk-like and interfacial spectra from such systems. With such a direct measure of the bulk line-shape, a comparison between the isolated DL spectrum and those from the pure air-water interface gives good indication of how significant the quadrupolar signal is and, more importantly, allows for it to be subtracted from the total line-shape.

The measured spectra for interfaces with both positively and negatively charged surfactants are shown in Figure 2b. For this, we use the surfactants dihexadeyldimethylammonium bromide (DHAB) and dihexadecyl phosphate (DHP), which are both highly insoluble (thus allowing for high surface coverage without substantial bulk concentrations) and free from any resonances that could interfere with the water bending response. On initial inspection, the spectra from both charges exhibit positive bands. The absence of a clear sign flip in the two responses immediately shows that the bulk quadrupolar signal must still be a significant contribution to the spectra. However, further comparison shows that, while the peaks in the spectra do not have opposite signs, the positive surfactant yields spectra that are enhanced in amplitude and slightly red-shifted (compared to pure air-water, see dashed black line) and the negative surfactant yields spectra that have diminished amplitudes and slight blue-shifts. This indicates that the spectra contain both a charge-dependent (dipolar) and charge-independent (quadrupolar) contribution, as previously reported[33].

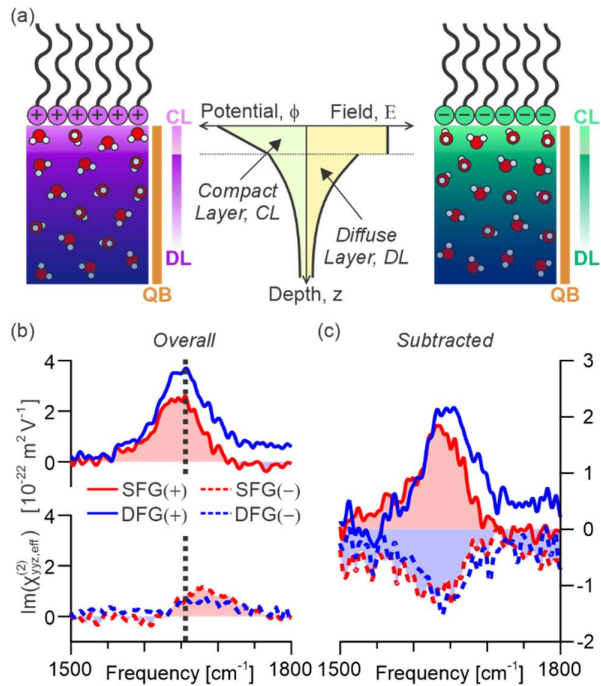

*Figure 2: (a) Schematic representation of the Gouy-Chapmann-Stern (GCS) model for charged aqueous interfaces, showing the evolution of electrostatic potential and electric field with depth, indicating the two distinct dielectric regions, the compact layer, CL, and the diffuse layer, DL. Also shown are schematics of the anisotropic water structure for both positive and negative charged surfaces as well as bar-representations of the different contributions to the SFG (and DFG) responses (including the bulk quadrupolar contribution, QB). (b) SFG and DFG spectra for the positive (solid) and negative (dashed) surfactants, with the dashed black line indicating the position of the peak frequency in the pure air-water spectrum. (c) The same surfactant spectra shown in (b) having subtracted the pure air-water interface spectrum shown in Figure 1d.*

To isolate the charge-dependent dipolar contribution to the spectra, one can effectively remove the quadrupolar contribution by subtracting the spectra from the pure air-water interface, with the results shown in Figure 2c. As mentioned above, the quadrupolar contribution should be the same in both systems and thus it should be removed from such a subtraction. This leaves only the dipolar contributions, which should clearly be dominated by the charged interfaces. It is clear from comparing these spectra in Figure 2c that they now display the expected sign-flip upon charge inversion, confirming their dipolar origin. Beyond this, comparing the corresponding SFG and DFG spectra shows that they are clearly phase-shifted, presenting different line-shapes. This highlights the significant anisotropic depth in these systems, with the DL contributions extending over 100 nm away from the interface (given a salinity of $< 10^{-5}$ M).

Even after removal of the quadrupolar signals the obtained spectra in Figure 2c still contain two distinct dipolar contributions from the charged interfaces, the signals from the CL and the DL, as well as potentially a minor contribution from the dipolar air-water spectrum introduced from the subtraction. To extract the pure bulk-like DL spectrum, we instead combine all four charged interface spectra (overall, not subtracted, Figure 2b). Full details of this extraction are given in Supplementary Information. Briefly, the difference spectrum between SFG and DFG for each charge removes their CL contributions, and then the subtraction of these difference spectra for the two charges eliminates the quadrupolar contributions while constructively combining the DL spectra (which have opposite signs for the two charges).

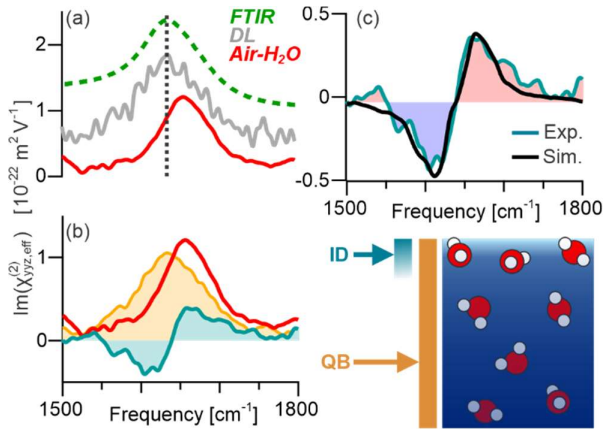

*Figure 3: (a) Spectra for the pure air-water interface (SFG, red) and diffuse layer (DL, grey) from the charged interfaces, extracted based on the GCS model, along with the linear FTIR spectrum (dashed green) for comparison. The three spectra have been offset for clarity. (b) The pure air-water interface SFG spectrum (red) along with its decomposition into an isotropic (bulk) quadrupolar contribution (orange) and an anisotropic (interfacial) dipolar contribution (blue). The spatial origins of these two sources are also schematically indicated. (c) The extracted dipolar contribution to the pure air-water interface spectrum overlapping with that obtained through MD simulations.*

The obtained DL response is shown in Figure 3a (grey trace), having also removed any influence of depth on its phase. The spectrum is shown alongside the pure air-water interface SFG spectrum (red, also shown in Figure 1a/d) and the FTIR spectrum of pure water, representing an approximation for the effective bulk $\chi^{(1)}$ response (dashed green). On comparison, the DL spectrum matches exceptionally well with the FTIR spectrum, with both presenting a single band centred at ~1637 cm$^{-1}$ with a FWHM of ~90 cm$^{-1}$. The only noticeable difference exists for the low-frequencies which are known to contain additional intensity in FTIR spectra owing to the first overtone of the low-frequency libration modes[45]. This agreement aligns perfectly with expectation as the DL signals probe bulk-like water, with any influence on the surface chemistry of the surfactants only affecting the CL. By contrast, if we compare the spectrum from the pure air-water interface to the bulk-like spectra of either the DL or FTIR, there are significant line-shape differences, particularly in peak frequency (1656 cf. 1637 cm$^{-1}$). This discrepancy hence must arise from the interfacial dipolar response in the pure air-water spectrum. As shown in the decomposition in Figure 3b, subtraction of the bulk line-shape (DL spectrum) from the pure air-water interface uncovers a residual component with a defined 'dip-peak' spectrum (see Supplementary Information for more detail on this subtraction). A comparison of this component to the ID-only prediction from MD simulations (Figure 3c) shows perfect agreement in both line-shape and absolute amplitude, therefore conclusively confirming that this spectral feature is indeed the elusive ID contribution.

*Elucidating the Depth-Dependent Interfacial Water Structure*

With the interfacial (ID) response isolated and the theoretical predictions experimentally verified, we can now confidently analyse the structural implications of the obtained spectrum. The 'dip-peak' line-shape is characteristic of two overlapping contributions: a red-shifted dip and a blue-shifted peak. Each of these spectral features then contains two important pieces of contrasting information, their peak frequency and their sign, which relate to the intermolecular connectivity and molecular orientation, respectively. To analyse the peak frequency, one typically compares the known frequency shifts due to H-bonding in bulk water, where a red-shift in the bending mode indicates weaker H-bonding (in direct contrast to the O-H stretch). For the sign of the response, one typically compares them to the sign obtained for the dipolar contribution from charged interfaces. As these responses arise from field-induced reorientation, the peak observed for the positively charged interface should be dominated by molecules with their dipoles pointing 'down' (towards bulk water) and the dip observed for the negatively charged interface from molecules with their dipoles pointing 'up' (towards the air phase), as indicated in Figure 4b. This would therefore suggest that the red-shifted dip for air-water arises from 'pointing up' molecules with weaker H-bonding, and the blue-shifted peak from 'pointing down' molecules that are more strongly interconnected.[46–48] This interpretation is fully compatible with the expectations of the interface as one would expect molecules 'pointing down' to have greater access to H-bonding than those pointing towards air.

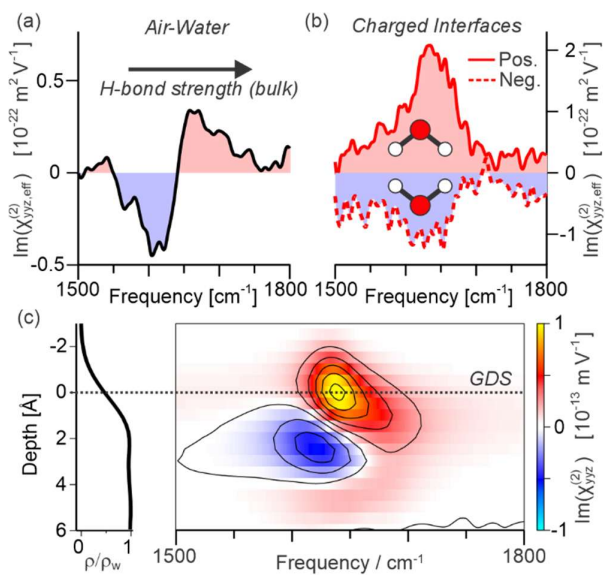

*Figure 4: (a) Extracted interfacial (dipolar) spectrum for the pure air-water interface, as shown in Figure 3c. (b) SFG spectra for both surfactants having subtracted the pure air-water interface spectra, as shown in Figure 2c, indicating the average preference for the molecular dipole pointing down (positive surfactants) and up (negative surfactants). (c) Depth-dependent ID-only bending response calculated from MD simulations, also showing the depth-dependent density profile and the Gibbs dividing surface (GDS), defining zero depth.*

To further analyse the interfacial structure, one can then turn to the depth-dependence of the bending response. This is shown in Figure 4c, which plots the 2D ID-only second-order susceptibility, $\chi^{(2)}_{yyz}(z)$, calculated from the MD simulations. From this, one can clearly identify both the positive response at higher frequencies and the negative response at lower frequencies. These appear, however, to be spatially separated, i.e. arising predominantly from different depths. Particularly, the blue-shifted positive contribution clearly sources from closer to the air-phase. Following the interpretation of the spectral line-shape discussed above, this would therefore suggest that the topmost water molecules are predominantly 'pointing down' and more strongly H-bonded than those located below which predominantly 'point up'. This picture clearly no longer fits with our expectations of the topmost water having dangling bonds and being more weakly H-bonded, or equally with previous analysis of the SFG stretching mode spectra[37,38,49–51], or even calculations of the preferential molecular orientation at the interface[52].

The above analysis of both the experimental spectra and its calculated depth-dependence clearly shows that the typical way of interpreting the bending mode spectrum must be incorrect and should be substantially revised. This can be done partially from first principles by considering the amplitude of the SFG response from an individual water molecule as a function of its molecular orientation (i.e., neglecting any frequency-dependence associated with its interconnectivity). The spatial mapping of the molecular coordinates onto the interface is described by three angles: tilt, twist, and azimuth. Here, the tilt angle is defined as the angle between the molecular dipole and the surface normal, the twist angle is the amount of rotation around the dipole axis, and the azimuth is simply associated with the direction of the in-plane projection of the molecular dipole, which is isotropically distributed at such liquid interfaces. A schematic of the tilt and twist angles is shown in Figure 5a. By combining this orientational projection with the molecular symmetry properties and a simple geometric description of the bending vibration (see Supplementary Information for details), Figure 5b then shows the calculated relative SFG amplitude of a single water molecule as a function of both the tilt and twist angles. For better clarity, Figure 5c gives schematics of the molecular orientation at the interface for selected combinations of the two angles.

The theoretical plot of SFG amplitudes in Figure 5b shows that both the magnitude- and sign-dependence of the bending response for different molecular orientations is far from simple. Specifically, while it is loosely true that molecules 'pointing up' (with tilt angles < 90°) yield negative responses and molecules 'pointing down' (with tilt angles >90°) yield positive responses, there are clear regions in the plot that show the opposite, with this highlighted by a contour line at zero amplitude (shown in white). Importantly, molecules 'pointing up' can yield positive or negative responses, depending on their twist angle. This hence raises the question whether the aforementioned contradiction between the depth-dependent response and the conventional interpretation of the spectral line-shape arises due to the observed signals sourcing from molecules with orientations in these sign-flip regions. In other words, is the twist angle at the interface orientationally restricted and correlated to the tilt angle?

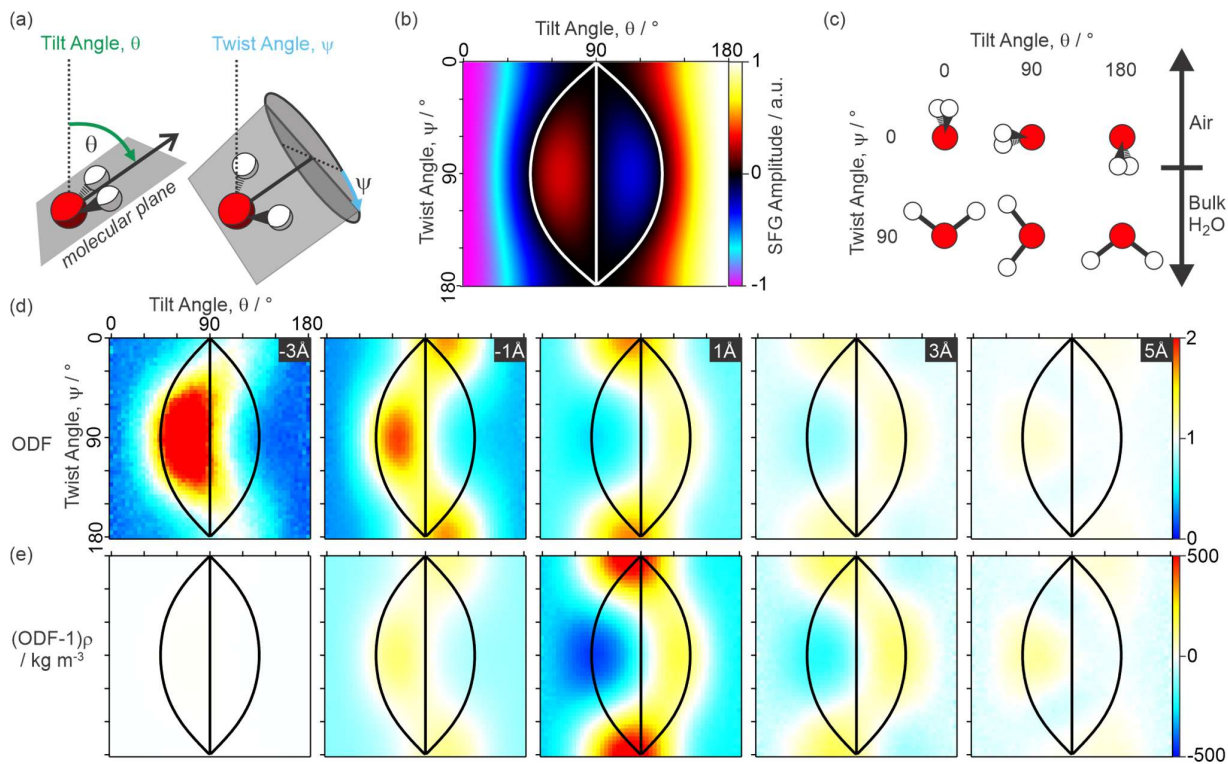

*Figure 5: (a) Schematic of the molecular tilt and twist angles, showing the defined molecular plane that contains all three atoms in $H_2O$. (b) Calculated SFG amplitudes for the bending mode (see Supplementary Information for details). (c) Representation of the molecular orientation at the interface for specific combinations of tilt and twist. (d) Calculated 3D orientational distribution function (tilt vs. twist vs. depth) from MD simulations, shown at specific depths through the interface, also shown in (e) having normalised for the molecular density across the interface.*

The answer to this question lies in the full 3D orientational distribution function (ODF) at the interface, i.e. tilt vs. twist vs. depth.[53] This was hence calculated from the MD simulations and is presented in Figure 5d. Specifically, Figure 5d shows the tilt vs. twist distribution at 2Å steps through the interface, where regions displaying yellow or red (i.e., ODF values >1) indicate molecular orientations that are populated more than they would be for an isotropic distribution, and those in cyan and blue (ODF values <1) represent the less populated orientations with respect to an isotropic distribution. On initial inspection, the ODF above the interface (towards the air phase), at -3Å, does indeed show that the preferential molecular orientation at this depth lies almost solely in the calculated sign-flip region (indicated by the same contour lines shown in white in Figure 5b being repeated here in black). Specifically, it shows a clear preference for water molecules with their dipoles pointing slightly 'up', but with a 90° twist, meaning they are effectively tilted in the molecular plane (see Figure 5a and c). Below this, at -1Å, the same molecular orientation is still preferential, but with a diminished enhancement, and a second orientation becomes favourable, specifically molecules pointing 'down' with 0°/180° twist. Thereafter, on the other side of the Gibbs diving surface (GDS, which defines z=0), this is then essentially mirrored for +1 and +3Å, with the preference being for 'pointing up' with 0° twist and 'pointing down' with 90° twist. Finally, for +5Å it appears to flip again, showing a similar distribution as for -1Å, but with significantly lower population enhancement due to the increasing tendency towards an isotropic distribution. Overall, this clearly demonstrates that the twist angle is not only highly restricted at the interface, but also strongly correlated to the tilt angle, and that both are strongly varying as a function of depth.

This depth-dependent flipping in the preferential molecular orientation (both tilt and twist angles) is strongly indicative of a dominant role of the local H-bond network on the orientational distribution. This is suggested as a 90° change in twist corresponds precisely to flipping between the H-bond donor plane (same as molecular plane, see Figure 5a) and the plane of the H-bond acceptors (O lone pairs). Therefore, it seems that the highly directional and approximately tetrahedrally oriented nature of the H-bonding is imprinted on the molecular orientations directly at the interface. While these macroscopic orientational correlations are not seen in the bulk due to its isotropic nature, they clearly appear at the interface. The resulting pronounced layering must therefore originate from a specific preferential orientation in the top-most layer that is induced by the macroscopic phase boundary. The orientation in this top-most layer then restricts the orientational distribution in the subsequent layers due to their thermodynamic preference for maximising their interconnectivity.

It is important to note that, while the regions towards the air-phase present the strongest preferential orientation, the density of water molecules closer to the air-phase is much smaller than on the other side of the GDS. This means that, considering their density, these regions contribute essentially nothing to the overall SFG signals. This effect is accounted for in Figure 5e which scales the ODF for density (but now with a different normalisation such that 0 corresponds to an isotropic distribution). As expected, the density-scaled ODF at -3Å shows no significant contributions. Below this, however, there are clear contributions from the above-mentioned orientations spanning the presented ~6-8Å range in depth.

As discussed above, analysis of the ODF shows that there are essentially four distinct classes of preferential orientations at the air-water interface: (1) pointing slightly 'up' with 90° twist, (2) pointing slightly 'down' with 0° twist, (3) pointing slightly 'up' with 0° twist, and (4) pointing slightly 'down' with 90° twist. As such, the anisotropic structure in terms of molecular orientation can be well-described from the depth-dependent populations of these four structural motifs, which are shown in Figure 6a through their depth-dependent contributions to the density-scaled ODF. The fill-colour of these traces approximately represents their contributed SFG amplitude, estimated from the calculated response shown in Figure 5b (where diminished populations compared to an isotropic distribution are assigned the opposite sign response). Figure 6b then shows a schematic representation of this anisotropic structure at the air-water interface, again where the fill-colour of the molecules highlights their contributed SFG amplitude. These approximate amplitudes (as well as those in Figure 6a) can be compared to the depth-dependent $\chi^{(2)}(z)$ calculated from simulations shown in Figure 6c (just as in Figure 4c). This comparison shows overall excellent agreement in terms of the depth-dependent sign of the response, with structural motifs (1) and (2) dominating the top-most region and both yielding a positive response, and motifs (3) and (4) dominating the region below, both yielding negative signals. Finally, the weak positive contribution from ~3-5Å arises from a slight preference for motifs (1) and (2) again.

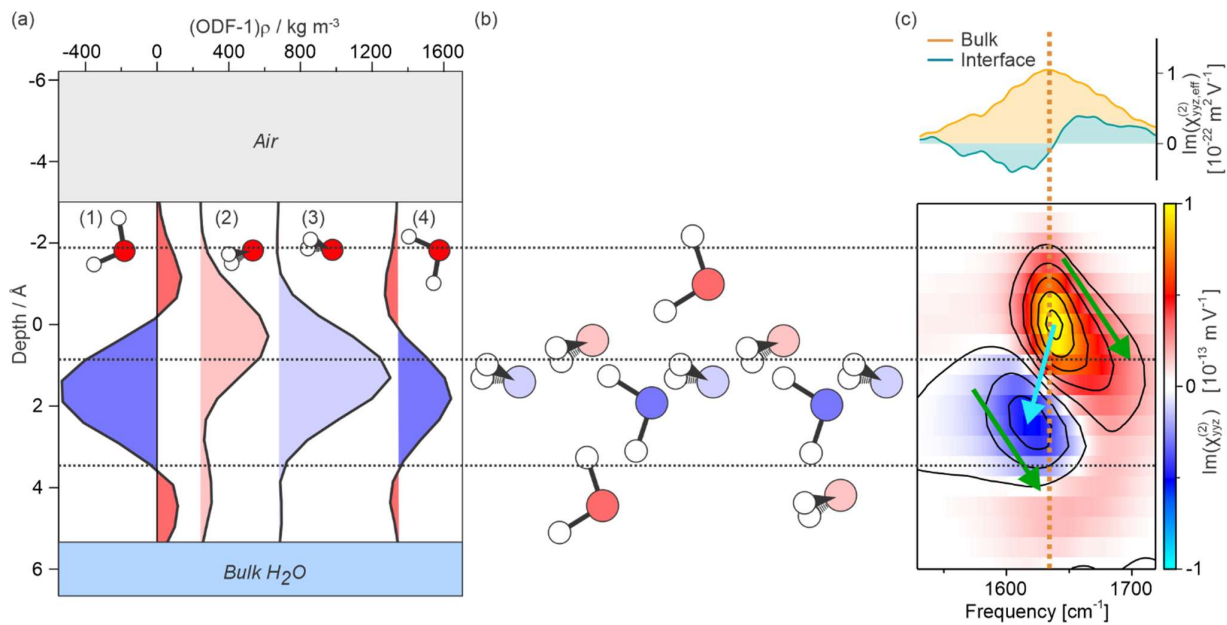

*Figure 6: (a) Depth-dependence of specific orientations from the density-normalised orientational distribution function. (b) Schematic representation of the preferential (anisotropic) orientation at the air-water interface. The fill-colours of the traces in (a) and molecules in (b) approximately represent the amplitude of their SFG contribution. (c) Depth-dependent second-order susceptibility, $\chi^{(2)}_{yyz}(z)$, shown with the experimentally determined interfacial spectrum, as well as the bulk-like spectrum for comparison. The green arrows highlight the depth-dependent blue-shift for each of the positive and negative responses, while the blue arrow highlights the red-shift between the different sign responses.*

At first sight, the 3D 'picture' of the anisotropic interfacial orientations in Figure 6b indicates that the structural motifs (2) and (3) which dominate the interfacial anisotropy from ~0-2Å could be part of a macroscopic 'buckled' 2D water network, as previously suggested from simulations.[52,54,55] However, for a clear assessment of any network structure, the precise intermolecular connectivity would have to be determined. Such an analysis is highly involved and thus requires substantial further work. Nevertheless, some insights can be obtained through the apparent depth-dependent frequency shifts in the bending response. On inspection of Figure 6c, there are two clear types of depth-dependent frequency shifts: (i) a distinct blue-shift of both features indicated by the green arrows, and (ii) a general red-shift of the negative feature cf. the positive feature, as indicated by the blue arrow. While the gradual depth-dependent blue-shift in both features could be a manifestation of the increasing H-bonding strength with depth, the overall displacement of the negative response to lower frequencies, despite its larger depth, strongly deviates from this. The different shifts of the bend and stretch water bands when going from vapor to the liquid phase (red for stretch and blue for bend) was recently explained in terms of the competition of the bond elongation when entering the liquid phase, which induces a red-shift, and frequency-dependent friction effects, which induce a blue-shift.[56] A similar competition of bond elongation and dissipation is expected when moving through the interface from vapor to bulk liquid. Interestingly, a comparison of the absolute frequencies of these features to the frequency of the bulk response (indicated by the dashed orange line in Figure 6c) actually shows that top-most layer seems to display a more bulk-like spectrum than the layers beneath. Overall, therefore, analysis of the depth-dependence of the vibrational frequency shows that the anisotropic interconnectivity at the air-water interface and its connection with the vibrational bending frequency is far from simple and thus requires further theoretical investigation. The above observations, however, highlight that either the interconnectivity is far from a monotonic function with depth, and/or that the factors governing the bending response at the interface, i.e. the different aspects of the H-bonding network, vastly differ from the bulk.

## Conclusions

In summary, by applying our recently developed depth-resolved second-order spectroscopy to study the H-O-H bending mode we have separated out a significant bulk-like quadrupolar contribution to the air-water interface spectrum and extracted the previously unobserved interfacial dipolar response. This is shown to quantitatively match the calculated spectrum from MD simulations, thus finally verifying the theoretical predictions and allowing a more in-depth investigation into the interfacial structure. Through an analysis of the spectral line-shape of the extracted interfacial response, as well as its calculated depth dependence, we then demonstrate that the conventional understanding and interpretation of the water spectrum in terms of 'pointing up' and 'pointing down' molecular dipoles is not only incorrect but misses out on a crucial part of the preferential orientation at the interface, namely the twist angle. From the full calculated orientational distribution, we clearly demonstrate that both the tilt and twist are highly restricted, correlated to one another, and show a pronounced layer / depth dependence. Specifically, we highlight that the anisotropic structure and its resulting SFG amplitudes can be well-described by four distinct classes of molecular orientation, with dipoles pointing either slightly 'up' or slightly 'down', and twist angles close to either 0 or 90°. This gives a new and detailed picture of the structural motifs that dominate and define the anisotropic structure of the air-water interface. Finally, by analysing the depth-dependence of the vibrational frequencies, we show that the interconnectivity at the interface is highly complex and either not well-described by a simple combination of orientation- and monotonic depth-dependence to the H-bonding, or that it shows a non-bulk-like connection between vibrational frequency and interconnectivity. In either case, this highlights the anomalous nature of the interfacial H-bond network.

## Acknowledgements

The authors thank V. Balos for his contribution to the design and implementation of the SFG/DFG spectrometer used in this work and acknowledge the Deutsche Forschungsgemeinschaft (DFG) for funding (Project-ID 221545957 - SFB 1078/C1).

## Competing Interests

There are no competing interests.

## Author Contributions

A.P.F., R. R. N, and M.T. conceived the project and designed the experiments. A.P.F. performed the experiments and L.L. undertook the simulations. Analysis of the data was performed by A.P.F. and L. L., while all authors discussed the interpretation of the results. A.P.F. drafted the manuscript which was edited by all authors. R.R.N., M.W., and M.T. supervised the work and acquired funding.

## Data Availability

Raw data will be made available upon reasonable request by contacting the corresponding author.

# Supplementary Information

The Importance of Layer-Dependent Molecular Twisting for the Structural Anisotropy of Interfacial Water

Alexander P. Fellows[1†], Louis Lehmann[2†], Álvaro Díaz Duque[1], Martin Wolf[1], Roland R. Netz[2], and Martin Thämer[1]*

[1] Fritz-Haber-Institut der Max-Planck-Gesellschaft, Faradayweg 4-6, 14195, Berlin, Germany

[2] Department of Physics, Freie Universität Berlin, Arnimallee 14, 14195, Berlin, Germany

† Equal contribution

* Corresponding author

thaemer@fhi-berlin.mpg.de

(tel.): +49 (0)30 8413 5220

# Experimental Methods

*Sample Preparation*

The spectroscopic measurements of the aqueous interfaces were performed on both $H_2O$ (Milli-Q, 18.2 MΩ·cm, <3 ppb TOC) and $D_2O$ (VWR Chemicals, 99.9% D) contained in custom-made PTFE troughs that were cleaned overnight with Piranha solution (3:1 sulfuric acid to 30% hydrogen peroxide solution) and subsequently thoroughly rinsed with ultrapure water. *Warning: Piranha solution is highly corrosive and an extremely powerful oxidizer. Great care must be taken with its preparation and use.*

For the measurements at charged interfaces, the two surfactants, dihexadecyl phosphate (DHP and dihexadeyldimethylammonium bromide (DHAB), were prepared as a 1 mg ml$^{-1}$ solution in chloroform and deposited dropwise onto the water surface in the PTFE troughs until surface saturation was reached. This procedure ensured a consistent well-packed monolayer film of sufficient density as to minimize any effects from Bénard-Marangoni convection.[1]

*Spectral Acquisition*

The details of the SFG and DFG interferometer used to measure the spectra shown here can be found elsewhere.[2] Briefly, the 7W 1kHz 800nm output from a Ti:sapphire laser (Astrella, Coherent) is fed into two independent optical parametric amplifiers (TOPAS Prime, Light Conversion), the first being used to generate mid-IR through DFG, and the second producing a signal beam that is subsequently frequency-doubled to produce a tunable visible upconversion. The IR output is split, with one part being collinearly overlapped with the visible in z-cut quartz to generate local oscillator (LO) references that are (along with the visible) subsequently collinearly overlapped with the second part. This generates a single collinear beam that is directed to the sample at 70° incidence angle to produce the signal SFG and DFG. The reflected beam is then filtered and collected on single channel detectors, implementing balanced detection. Furthermore, for greater phase accuracy, the sample scans are recorded alongside a z-cut quartz reference through shot-to-shot referencing.

The acquired spectra were recorded in the time domain from -400 to 5000 fs in 0.8 fs steps to ensure sufficient spectral resolution. The presented spectra represent co-averages across multiple individual scans. Specifically, the pure air-water interface represents the average of 62000 spectra, the DHAB measurements across 73000 spectra, and the DHP across 79000 spectra. During the measurements, the entire optical path was purged with dry, $CO_2$-scrubbed air to minimize atmospheric absorption. Additionally, to account for any changes in optical path due to sample evaporation, the sample height was corrected during each measurement using an automated stage.

*Data Treatment*

To correct the spectra for amplitude and phase, they were referenced using a transfer function measurement of z-cut quartz recorded from -400 to 3000 fs in 0.8 fs steps, taking the response from quartz to have a phase of ±90° as a close approximation, neglecting any surface effects from the quartz[3]. For a more accurate phase correction, the spectra were further corrected using the SFG and DFG responses of the carbonyl stretch from a dipalmitoylphosphatidylcholine monolayer on fused silica (averaged over 18000 scans), which is highly surface localized and thus should result in a minute phase difference.[4–6] The resulting amplitude- and phase-corrected spectra were then corrected for the non-resonant response by subtracting analogous measurements from $D_2O$.

# Calculation of Spatially Resolved $\chi^{(2)}$ from Molecular Dynamics Simulations

The theoretical prediction of the spatially resolved electric dipolar contribution to the second-order susceptibility was calculated from *ab initio*-parameterized molecular dynamics simulations, full details of which can be found elsewhere.[6] From this, the 2D frequency vs. depth map of the yyz component, $\chi^{(2)}_{yyz}(z)$, was extracted in absolute units. The corresponding spectrum of the effective second-order susceptibility, $\chi^{(2)}_{yyz,eff}$, was then calculated by integrating the response over depth and applying the corresponding depth-related phase-shift using the wavevectors taken from the experimental settings.

For the analysis of the depth-dependence of the response, only the intramolecular component of $\chi^{(2)}_{yyz}(z)$ was used as it has a significantly better signal-to-noise ratio and can be better understood when assessing the anisotropic structure at the interface. Particularly, for comparison to the SFG amplitudes of the bending response for different molecular orientations calculated purely on theoretical grounds (shown in Figure 5b in the main text), using only the intramolecular response is a more direct comparison. In any case, the comparison between the purely intramolecular contribution to $\chi^{(2)}_{yyz}(z)$ and the total response presented in Figure S1 shows that neglecting the intermolecular component has little effect. The total response shows good qualitative overlap with the intramolecular response, reproducing the same positive and negative features in similar positions and still shows the main frequency-shifts as highted in Figure 6c and discussed in the main text. Any differences between the two plots in Figure S1 are within the uncertainty associated with the noise introduced by including the intermolecular contribution.

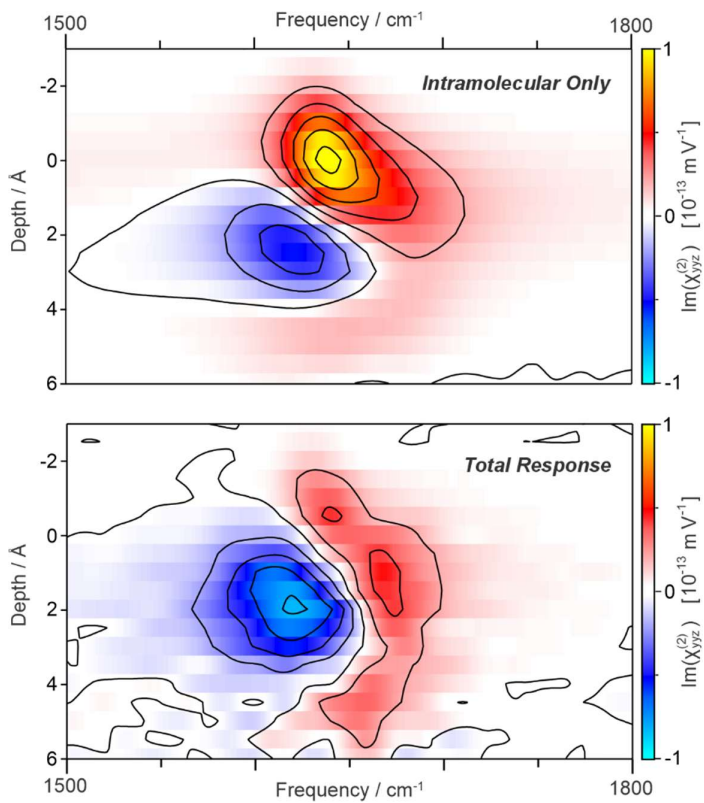

*Figure S1: Comparison of the intramolecular contribution to $\chi^{(2)}_{yyz}(z)$ and its total response.*

The simulations were performed for both $H_2O$ and $D_2O$, giving two independent comparisons of the resulting dipolar spectrum to experiment. While the comparison to the calculated $H_2O$ spectrum, in principle, directly compared the same frequency range, the minor nuclear quantum effects (smaller than those in the O-H stretch) that are not accounted for in the simulations result in the calculated spectrum being shifted by ~35 cm$^{-1}$ to higher frequencies. Nevertheless, shifting the spectrum by this value results in exceptional overlap, as shown in Figure S2.

When comparing the experimental $H_2O$ spectrum to the calculated $D_2O$ spectrum, one must account for the natural frequency shift caused by the change in reduced mass. This results in an effective frequency conversion factor of $\frac{3}{\sqrt{5}}$ which, when applied to the calculated spectrum, also yields exceptional overlap to the experimental results, as shown in Figure S2.

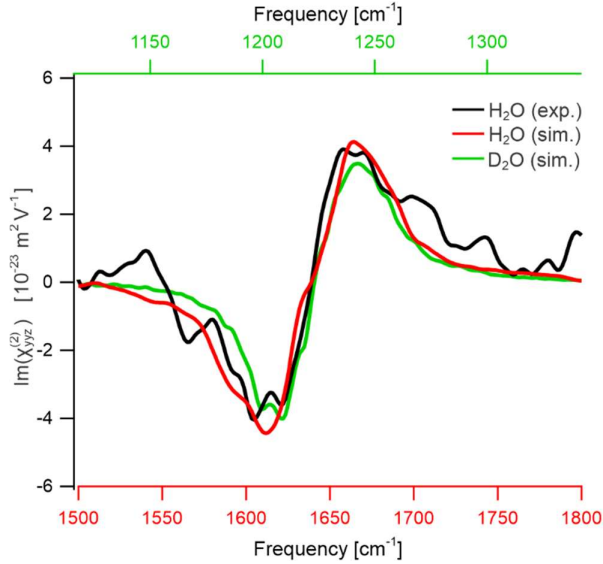

*Figure S2: Comparison between the experimentally isolated ID spectrum and the predicted ID-only spectra of both $H_2O$ and $D_2O$ calculated from simulations. The $D_2O$ spectrum is shown on the top axis, having calculated the corresponding frequency range based on the difference in reduced mass for the bending vibration.*

## Spectral Contributions, GCS Theory, and Extracting the DL Spectra

The general second-order response (SFG or DFG) from an interface probes the effective second-order susceptibility given by Eq. 1, containing both an electric dipolar (interfacial dipolar, ID) and quadrupolar (Q, taken here as combining electric quadrupolar and magnetic dipolar mechanisms, as per the conventions of Morita[7]) contribution.

$$\chi_{eff}^{(2)} = \chi_{ID}^{(2)} + \chi_Q^{(2)} \tag{1}$$

For the pure air-water interface, the dipolar response can be written as in Eq. 2, including its depth-dependent decay over the anisotropic decay length, z'.[4,5] Upon integration, a decay length-dependent amplitude-factor and phase-factor are introduced, as shown in Eq. 3. For very small decay lengths, as is the case for pure air-water[6], the amplitude-factor can be taken to be z' and the phase-factor, 1, leading to Eq. 4.

$$\chi^{(2)}_{eff,ID} = \int_0^\infty \chi^{(2)}_{ID} e^{\left(i\Delta k_z - \frac{1}{z'}\right)z} dz \tag{2}$$

$$= \frac{1}{\sqrt{\left(\frac{1}{z'}\right)^2 + \Delta k_z^2}} \chi^{(2)}_{ID} e^{i\,\text{atan}(\Delta k_z z)} \tag{3}$$

$$\approx z' \chi^{(2)}_{ID} \tag{4}$$

On the introduction of a charged surfactant at the interface, however, a static electric field is induced, meaning the ID contribution can be further broken down into a $\chi^{(2)}_D$ and $\chi^{(3)}_D$ response, as in Eq. 5, with the latter arising from depths extending much further away from the interface (ultimately controlled by the screening of the field). Based on the Gouy-Chapman-Stern (GCS) model, the ID response can be further broken down into a compact layer (CL) and diffuse layer (DL) contribution, as described by Eq. 6, with the latter containing only the field-induced term. By considering the CL to be much thinner than the coherence length and combining the $\chi^{(2)}_D$ and $\chi^{(3)}_D$ responses, the CL contribution can be represented by a single term with minimal propagation-induced phase-shift, $\chi^{(2)}_{CL,eff}$. Furthermore, applying the exponentially decaying potential in the DL allows the integral in Eq. 6 to be evaluated. Overall, this allows the ID response to be written as in Eq. 7, where $\Phi_S$ is the Stern potential.[8]

$$\chi^{(2)}_{ID} = \int_0^\infty \left(\chi^{(2)}_D + \chi^{(3)}_D E_{DC}\right) e^{i\Delta k_z z} dz \tag{5}$$

$$= \int_0^{z_S} \left(\chi^{(2)}_{D,CL} + \chi^{(3)}_{D,CL} E_{DC}\right) e^{i\Delta k_z z} dz + \int_{z_S}^\infty \chi^{(3)}_{D,DL} E_{DC} e^{i\Delta k_z z} dz \tag{6}$$

$$= \chi^{(2)}_{CL,eff} + \frac{1}{2} \Phi_S \chi^{(3)}_{DL} \cdot \left(1 + e^{2i\,\text{atan}(\Delta k_z z_{DL})}\right) \tag{7}$$

With this general description, the SFG and DFG responses for the two, oppositely charged surfactants (DHAB and DHP) can be written as in Eqs. 8-11.

$$\chi^{(2)}_{DHAB,SFG} = \chi^{(2)}_{CL,DHAB} + \frac{1}{2} \Phi_{DHAB} \chi^{(3)}_{DL}\left(1 + e^{2i\phi_{SFG}}\right) + \chi^{(2)}_{IQ,DHAB} + \chi^{(2)}_{IQB} + \chi^{(2)}_{QB,SFG} \tag{8}$$

$$\chi^{(2)}_{DHAB,DFG} = \chi^{(2)}_{CL,DHAB} + \frac{1}{2} \Phi_{DHAB} \chi^{(3)}_{DL}\left(1 + e^{2i\phi_{DFG}}\right) + \chi^{(2)}_{IQ,DHAB} + \chi^{(2)}_{IQB} + \chi^{(2)}_{QB,DFG} \tag{9}$$

$$\chi^{(2)}_{DHP,SFG} = \chi^{(2)}_{CL,DHP} + \frac{1}{2} \Phi_{DHP} \chi^{(3)}_{DL}\left(1 + e^{2i\phi_{SFG}}\right) + \chi^{(2)}_{IQ,DHP} + \chi^{(2)}_{IQB} + \chi^{(2)}_{QB,SFG} \tag{10}$$

$$\chi^{(2)}_{DHP,DFG} = \chi^{(2)}_{CL,DHP} + \frac{1}{2} \Phi_{DHP} \chi^{(3)}_{DL}\left(1 + e^{2i\phi_{DFG}}\right) + \chi^{(2)}_{IQ,DHP} + \chi^{(2)}_{IQB} + \chi^{(2)}_{QB,DFG} \tag{11}$$

In these descriptions, the general quadrupolar contribution is broken down into its three contributors, based on the conventions of Morita[7], with IQ representing the anisotropic interfacial response, IQB being the bulk-like interfacial contribution, and QB being the fully isotropic bulk response. Furthermore, any parameter that is dependent on the surfactant is given a unique label. This includes the CL susceptibility, $\chi_{CL}^{(2)}$, which contains all of the chemistry-driven structural alterations due to the direct solvation of the surfactant, the Stern potential that is modulated by the surface charge density and specific dielectric structure in the CL, and the IQ quadrupolar response that is also dependent on the molecular structure directly at the interface. Furthermore, as SFG and DFG have differing wavevector mismatches, they lead to unique propagation phases. While this effect is negligible for the highly surface-localised responses namely the CL, IQ and IQB, the phase factor in the DL contribution and overall QB contributions will be different for each response, as indicated by their labels.

Importantly, due to the sign-flip in charge of the two surfactants, $\chi_{CL}^{(2)}$ will have opposite signs for each, as will the two Stern potentials, $\Phi_{DHAB}$ and $\Phi_{DHP}$. This means the overall ID response will flip sign between the two surfactants, but it may well have slightly different line-shapes owing to the surface-chemistry contributions in the CL contribution and likely different surface charged densities. By contrast, all the quadrupolar responses will have the same sign for both surfactants, but potentially a slightly different line-shape for the IQ response.

With these descriptions in hand, it becomes clear that combining all four responses can retrieve the bulk-like DL spectrum, as shown in Eq. 12, having corrected for the propagation phase associated with the Debye length.

$$\frac{1}{2}(\Phi_{DHAB} - \Phi_{DHP})\chi_{DL}^{(3)} = \frac{\chi_{DHAB,DFG}^{(2)} - \chi_{DHAB,SFG}^{(2)} - \chi_{DHP,DFG}^{(2)} + \chi_{DHP,SFG}^{(2)}}{e^{2i\phi_{DFG}} - e^{2i\phi_{SFG}}} \qquad (12)$$

## Subtraction of Bulk-like Contribution from Air-Water

As mentioned in the main text, the desired ID contribution to the pure air-water interface spectra can be isolated by subtracting the bulk-like DL line-shape obtained using Eq. 12 as it should have an identical line-shape to the bulk-like quadrupolar contribution. This, however, requires knowledge of the correct amount of the DL spectrum to subtract i.e., the absolute amplitude of the bulk-like quadrupolar contribution (which will generally be different to the DL contribution). As this contribution is also present for the charged interface responses, its amplitude can be estimated from these spectra. Specifically, the summation of the four spectra given by Eqs. 8-11 should largely cancel the dipolar responses, only leaving residues due to their differences in specific chemistry-driven CL structure and Stern potentials. This means the total response should be dominated by the quadrupolar contributions that, as indicated in the main text, represent a significant part of the overall response. Therefore, by neglecting these slight dipolar residuals and the IQ contributions, the amplitude of the two bulk-like quadrupolar responses, which should be unchanged between the charged interface spectra and those at the pure air-water interface, can be estimated, and the appropriate amount of the DL line-shape subtracted. This approach was used to isolate the ID spectrum presented in the main text.

For further validation of this approach, one can also compare the simulated spectra to the residual experimental spectrum having subtracted different amplitudes of the DL line-shape. Such spectra are shown below in Figure S2, with the original SFG spectrum in red and the subtracted spectra in black (with the thicker black spectrum indicating the extracted ID contribution using the amplitude estimated from the charged interface spectra). With different subtracted amplitudes, it is clear that the overall line-shape highly differs, with these differences manifesting as changes in the zero-crossing and both the amplitudes and frequencies of the apparent 'dip' and 'peak' within the spectrum (or loss of either component). Therefore, a comparison between the different subtracted spectra and the calculated ID spectrum from the MD simulations gives three independent parameters to assess the quality of the match. On inspection of Figure S2 above as well as Figure 3 in the main text, the subtracted spectrum using the amplitude isolated from the charged interfaces yields exceptional overlap with all three of these parameters, far better than for any of the other presented subtractions in Figure S3. The quality of this match thus provides mutual validation of both the experimental and simulated spectra, suggesting that the desired experimental ID contribution is indeed isolated from this approach.

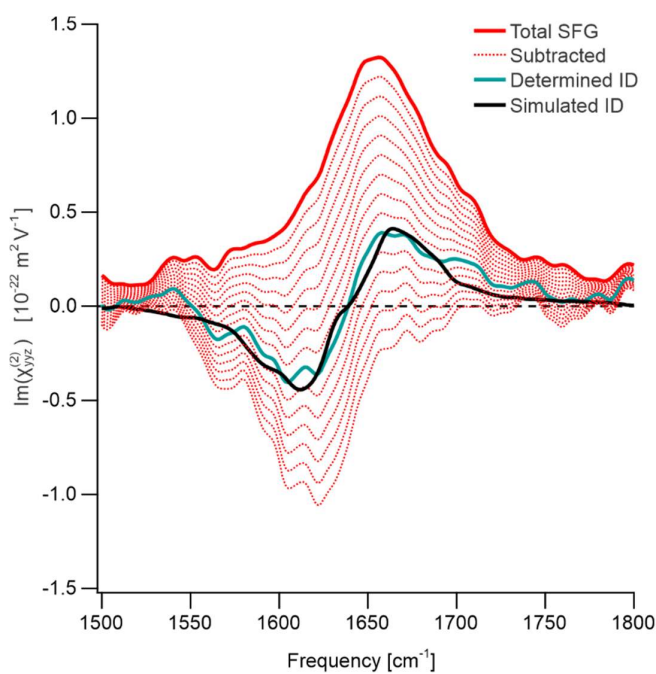

*Figure S3: SFG spectra of the pure air-water interface having subtracted different amounts of the obtained DL spectrum. The original SFG spectrum is shown in solid red, the spectra with different subtractions in dotted red, the determined ID line-shape using the subtraction amplitude estimated from the charged interface spectra shown in solid turquoise, and the calculated ID spectrum from MD simulations in solid black.*

# Theoretical Prediction of the H-O-H Bending Response

If we take the molecular symmetry of each $H_2O$ molecule as being $C_{2v}$, then there are 7 non-zero hyperpolarisability tensor elements: *aac*, *bbc*, *ccc*, *aca=caa*, and *bcb=cbb* (in the molecular frame coordinates (*a,b,c*) defined by the symmetry group, see Figure S4).

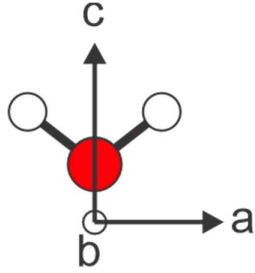

*Figure S4: Molecular-frame coordinate of water.*

For the H-O-H bending mode, which has the *A1* irreducible representation in $C_{2v}$, the relevant contributions are *aac*, *bbc*, and *ccc*, with the individual hyperpolarisability elements being given by Eq. 13, as the product of the partial derivatives of the Raman polarizability tensor, $\alpha^{(1)}$, and the IR transition dipole moment, $\mu$, evaluated in the equilibrium geometry, where $q$ represents the normal coordinate of the vibrational mode.[9]

$$\beta_{ijk} \propto \left(\frac{\partial \alpha_{ij}^{(1)}}{\partial q}\right)_0 \left(\frac{\partial \mu_k}{\partial q}\right)_0 \tag{13}$$

This shows that the three relevant components for the bending mode all involve the coupling of the IR along the dipole axis (*c*) and symmetric (diagonal) polarizabilities along each of the three molecular-frame coordinates.

The Euler transformation described by Eq. 14 can then be used to convert between the lab-frame and molecular-frame coordinates and isolate the *yyz* component of the second-order susceptibility.

$$\begin{pmatrix} x \\ y \\ z \end{pmatrix} = \begin{pmatrix} \cos\psi\cos\phi - \cos\theta\sin\phi\sin\psi & -\sin\psi\cos\phi - \cos\theta\sin\phi\cos\psi & \sin\theta\sin\phi \\ \cos\psi\sin\phi + \cos\theta\cos\phi\sin\psi & -\sin\psi\sin\phi + \cos\theta\cos\phi\cos\psi & -\sin\theta\cos\phi \\ \sin\theta\sin\psi & \sin\theta\cos\psi & \cos\theta \end{pmatrix} \begin{pmatrix} a \\ b \\ c \end{pmatrix} \tag{14}$$

If we then consider the in-plane Euler angle, $\phi$, to be isotropically distributed, the *yyz* susceptibility component is given by Eq. 15 (given that $\langle\sin^2\phi\rangle = \langle\cos^2\phi\rangle = \frac{1}{2}$ and $\langle\sin\phi\cos\phi\rangle = 0$).

$$\chi_{yyz}^{(2)}(A_1) = \frac{1}{2}N[(\cos^2\psi\,\beta_{aac} + \sin^2\psi\,\beta_{bbc} + \beta_{ccc})\cos\theta$$
$$+ (\sin^2\psi\,\beta_{aac} + \cos^2\psi\,\beta_{bbc} - \beta_{ccc})\cos^3\theta] \tag{15}$$

To simplify Eq. 15, we then consider the relations between the three hyperpolarizability components. We first assume that the transition polarizability is dominated by the rotation of the OH-bond orbitals.

Then, by denoting the OH bond polarizability tensor parallel to a OH-bond as $\alpha_\parallel$ and the perpendicular component as $\alpha_\perp$, and summing the polarizability contributions from the two OH-bond orbitals, the different components of the polarizability in the molecular frame can be expressed as in Eqs. 16-18.

$$\alpha_{aa}(q) = 2\sin^2\left(\frac{\tau}{2} + q\right)\alpha_\parallel + 2\cos^2\left(\frac{\tau}{2} + q\right)\alpha_\perp \tag{16}$$

$$\alpha_{bb}(q) = \alpha_\perp \tag{17}$$

$$\alpha_{cc}(q) = 2\cos^2\left(\frac{\tau}{2} + q\right)\alpha_\parallel + 2\sin^2\left(\frac{\tau}{2} + q\right)\alpha_\perp \tag{18}$$

In these expressions, the vibrational coordinate, $q$, for the bending mode is considered as simply an angular change to the dihedral angle of the two O-H bonds, $\tau \approx 104°$.

For the bending mode, we know that $\left(\frac{\partial \mu_i}{\partial q}\right)_0 = \delta_{ic}\Delta\mu$, where $\Delta\mu$ is the transition dipole moment. This leads to the expressions for the three components of the molecular hyperpolarizability given in Eqs. 19-21.

$$\beta_{aac} = \left(\frac{\partial \alpha_{aa}^{(1)}}{\partial q}\right)_0 \left(\frac{\partial \mu_c}{\partial q}\right)_0 = 2\sin\tau\,(\alpha_\parallel - \alpha_\perp)\Delta\mu \tag{19}$$

$$\beta_{bbc} = \left(\frac{\partial \alpha_{bb}^{(1)}}{\partial q}\right)_0 \left(\frac{\partial \mu_c}{\partial q}\right)_0 = 0 \cdot \Delta\mu \tag{20}$$

$$\beta_{ccc} = \left(\frac{\partial \alpha_{cc}^{(1)}}{\partial q}\right)_0 \left(\frac{\partial \mu_c}{\partial q}\right)_0 = -2\sin\tau\,(\alpha_\parallel - \alpha_\perp)\Delta\mu \tag{21}$$

With these approximations, Eq. 15 can be rewritten as in Eq. 22.

$$\begin{aligned}\chi^{(2)}_{yyz}(A_1) &= \frac{1}{2}N\beta_{ccc}\cos\theta\,[(1 - \cos^2\psi) \\ &\quad -(1 + \sin^2\psi)\cos^2\theta] \\ &= \frac{1}{2}N\beta_{ccc}\cos\theta\,[\sin^2\theta\,(1 + \sin^2\psi) - 1]\end{aligned} \tag{22}$$

Figure 5b in the main text shows the calculation of the *yyz* component of the second-order susceptibility, relative to the *ccc* component of the molecular hyperpolarizability, as a function of the molecular tilt and twist angles, and includes a contour line corresponding to zero-values in white, indicating where the response changes sign.